\documentclass[11pt,reqno]{amsart}
\usepackage{t1enc}
\usepackage[latin1]{inputenc}
\usepackage{amsmath,latexsym,amssymb,graphicx,dsfont,amsthm,amsfonts}
\usepackage{color}
\usepackage{umoline}
\usepackage{float}
\usepackage{apalike}

\newcommand{\elo}[1]{\mathrm{Elo}_{#1}}

\begin{document}

\begin{center}
~\\[1cm]
\huge{UEFA EURO 2020 Forecast via Nested Zero-Inflated Generalized Poisson Regression}
\\[2cm]
\large{Lorenz A. Gilch}
\\[1cm]
\normalsize
Department of Informatics and Mathematics, University of
Passau\\\texttt{Lorenz.Gilch@Uni-Passau.de}\\[3cm]

\includegraphics[scale=0.4]{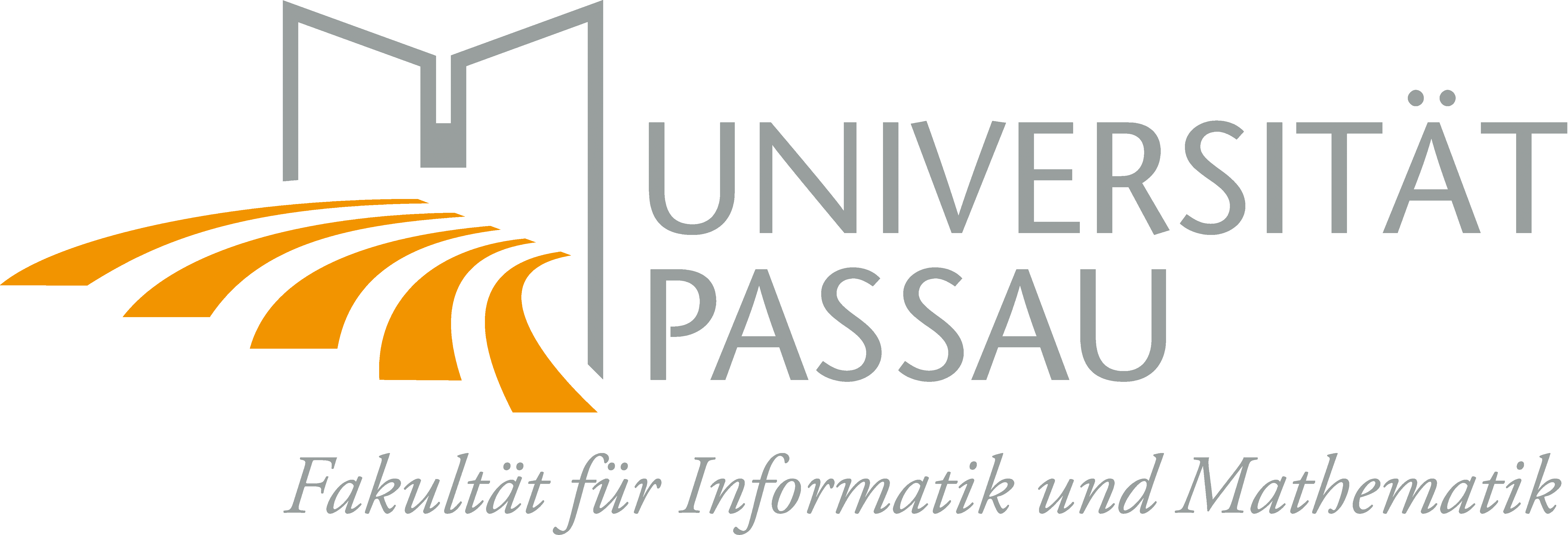} \\[1cm]
Technical Report, Number MIP-2101 \\
Department of Informatics and Mathematics\\
University of Passau, Germany\\
June 2021
\end{center} 

\thispagestyle{empty}
\newpage
\setcounter{page}{1}

\numberwithin{equation}{section}

\title[UEFA EURO 2020 Forecast via Nested ZIGP Regression]{UEFA EURO 2020 Forecast via Nested Zero-Inflated Generalized Poisson Regression}
\author{Lorenz A. Gilch}


\address{Lorenz A. Gilch: Universit\"at Passau, Innstrasse 33, 94032 Passau, Germany}

\email{Lorenz.Gilch@uni-passau.de}
\urladdr{http://www.math.tugraz.at/$\sim$gilch/}
\date{\today}

\maketitle

\begin{abstract}
This report is devoted to the forecast of the UEFA EURO 2020, Europe's continental football championship, taking place across Europe in June/July 2021.
We present the simulation results for this tournament, where the simulations
are based on a zero-inflated generalized Poisson regression model that includes the Elo points of the participating teams and the location of the matches as covariates and incorporates differences of team-specific skills. The proposed model allows predictions in terms of probabilities in order to quantify the chances for each team to reach a certain stage of the tournament. We use Monte Carlo simulations for estimating the outcome of each single match of the tournament, from which we are able to simulate the whole tournament itself. The model is fitted on all football games of the participating teams since 2014 weighted by date and importance. 
\end{abstract}

\section{Introduction}

Football is a typical low-scoring game and games are frequently decided through single events in the game. While several factors like extraordinary individual performances,  individual errors, injuries,  refereeing errors or just lucky coincidences are hard to forecast, each team has its strengths and weaknesses (e.g., defense and attack) and most of the results reflect the qualities of the teams. We follow this idea in order to derive probabilities 
for the exact result of a single match between two national teams, which involves the following four ingredients for both teams:
\begin{itemize}
\item Elo ranking
\item attack strength 
\item defense strength
\item location of the match
\end{itemize}
The  complexity of the tournament with billions of different outcomes makes  it very difficult to obtain accurate estimates of the probabilities of certain events. Therefore, we do \textit{not} aim on forecasting the exact outcome of the tournament, but we want to make the discrepancy between the participating teams \textit{quantifiable} and to measure the chances of each team to reach certain stages of the tournament or to win the cup. In particular, since the groups are already drawn and the tournament structure for each team (in particular, the way to the final) is set, the idea is to measure  whether a team has a rather simple or hard way to the final. 
\par
Since this is a technical report with the aim to present simulation results, we omit a detailed description of the state of the art and refer to \cite{gilch:afc19} and \cite{gilch-mueller:18} for a discussion of related research articles and a comparison to related models and covariates under consideration. 
\par
As a quantitative measure of the participating team strengths in this article,  we use the \textit{Elo ranking} (\texttt{http://en.wikipedia.org/wiki/World\_Football\_Elo\_Ratings}) instead of the FIFA ranking (which is a simplified Elo ranking since July 2018), since the calculation of the FIFA ranking changed over time and the Elo ranking is  more widely used in football forecast models. See also \cite{GaRo:16} for a discussion on this topic and a justification of the Elo ranking.  The model under consideration shows a good fit, the obtained forecasts are conclusive and give \textit{quantitative insights} in each team's chances.

\section{The model}
\label{sec:model}

\subsection{Preliminaries}
\label{subsec:goals}
The simulation in this article works as follows: each single match is modeled as $G_{A}$:$G_{B}$, where $G_{A}$ (resp.~$G_{B}$) is the number of goals scored by team A (resp.~by team B).  Each single match's exact result is forecasted, from which we are able to simulate the course of the whole tournament. Even the most probable tournament outcome has a probability very close to zero  to be actually realized. Hence, deviations of the true tournament outcome from the model's most probable one are not only possible, but most likely. However, simulations of the tournament yield estimates of the  probabilities for each team to reach certain stages of the tournament and allow to make the different team's chances \textit{quantifiable}. 
\par
We are interested to give quantitative insights into the following questions:
\begin{enumerate}
\item Which team has the best chances to become new European champion?
\item How big are the probabilities that a team will win its group or will be eliminated in the group stage?
\item How big is the probability that a team will reach a certain stage of the tournament?
\end{enumerate}

\subsection{Involved data}

The main idea is to predict the exact outcome of a single match based on a regression model which takes the following individual characteristics into account:
\begin{itemize}
\item Elo ranking of the teams
\item Attack and defense strengths of the teams
\item Location of the match (either one team plays at home or the match takes place on neutral ground)
\end{itemize}
We use an Elo rating system, see  \cite{Elo:78}, which includes modifications to take various football-specific variables (like home advantage, goal difference, etc.) into account. The Elo ranking is published by the website \texttt{eloratings.net}, from where also all historic match data was retrieved.
\par
We give a quick introduction to the formula for the Elo ratings, which uses the typical form as described in \texttt{http://en.wikipedia.org/wiki/World\_Football\_Elo\_Ratings}: let $\mathrm{Elo}_{\mathrm{before}}$ be the Elo points of a team before a match; then the Elo points $\mathrm{Elo}_{\mathrm{after}}$  after the match against an opponent with Elo points $\mathrm{Elo}_{\mathrm{Opp}}$ is calculated as follows:
$$
\mathrm{Elo}_{\mathrm{after}}= \mathrm{Elo}_{\mathrm{before}} + K\cdot G\cdot (W-W_e),
$$
where 
\begin{itemize}
\item $K$ is a weight index regarding the tournament of the match (World Cup matches have wight $60$, while continental tournaments have weight $50$)
\item $G$ is a number taking into account the goal difference:
$$
G=\begin{cases}
1, & \textrm{if the match is a draw or won by one goal,}\\
\frac{3}{2}, & \textrm{if the match is won by two goals,}\\
\frac{11+N}{8}, & \textrm{where $N$ is the goal difference otherwise.}
\end{cases}
$$
\item $W$ is the result of the match: $1$ for a win, $0.5$ for a draw, and $0$ for a defeat.
\item $W_e$ is  the expected outcome of the match calculated as follows:
$$
W_e=\frac{1}{10^{-\frac{D}{400}}+1},
$$
where $D=\mathrm{Elo}_{\mathrm{before}}-\mathrm{Elo}_{\mathrm{Opp}}$ is the difference of the Elo points of both teams.
\end{itemize}
The Elo ratings  on 8 June 2021  for the top $5$ participating nations in the UEFA EURO 2020 (in this rating) were as follows:
\begin{center}
\begin{tabular}{|c|c|c|c|c|}
\hline
Belgium & France & Portugal & Spain & Italy \cr
\hline
2100 & 	2087    & 	2037 & 2033	& 2013 \cr
\hline
\end{tabular}
\end{center}

The forecast  of the exact result of a match between teams $A$ and $B$ is modelled as 
$$
G_A\ : \ G_B,
$$
where $G_A$ and $G_{B}$ are the numbers of goals scored by team $A$ and $B$.
The  model is   based on  a \textit{Zero-Inflated Generalized Poisson} (ZIGP) regression model, where we assume  $(G_{A}, G_{B})$ to be a bivariate zero-inflated generalized Poisson distributed random variable. The distribution of $(G_{A}, G_{B})$   
will depend on the current Elo ranking $\elo{A}$ of team $A$, the Elo ranking $\elo{B}$ of team $B$ and the location of the match (that is, one team either plays at home or the match is taking place on neutral playground). The model is  fitted  
using  all matches of the participating teams  between 1 January 2014 and 7 June 2021. The historic match data is weighted according to the following criteria:
\begin{itemize}
\item Importance of the match
\item Time depreciation 
\end{itemize}
In order to weigh the historic match data for the regression model we use the following date weight function for a match $m$:
$$
w_{\textrm{date}}(m)=\Bigl(\frac12\Bigr)^{\frac{D(m)}{H}},
$$
where $D(m)$ is the number of days ago when the match $m$ was played and $H$ is the half period in days, that is, a match played $H$ days ago has half the weight of a match played today. Here, we choose the half period as $H=365\cdot 3= 3\,\textrm{years}$ days; compare with Ley, Van de Wiele and Hans Van Eetvelde \cite{Ley:19}.
\par
For weighing the importance of a match $m$, we use the match importance ratio in the FIFA ranking which is given by
$$
w_{\textrm{importance}}(m)=\begin{cases}
4, & \textrm{if $m$ is a World Cup match},\\
3, & \textrm{if $m$ is a continental championship/Confederation Cup match},\\
2.5, & \textrm{if $m$ is a World Cup or EURO qualifier/Nations League match},\\
1, & \textrm{otherwise}.\\
\end{cases}
$$ 
The overall importance of a single match from the past will be assigned as
$$
w(m)= w_{\textrm{date}}(m) \cdot w_{\textrm{importance}}(m).
$$
In the following subsection we explain the model for forecasting a single match, which in turn is used for simulating the whole tournament  and  determining the likelihood of the success for each participant.

\subsection{Nested Zero-Inflated Generalized Poisson Regression}

We present a \textit{dependent} Zero-Inflated Generalized Poisson regression approach for estimating the probabilities of the exact result of single matches. Consider a match between two teams $A$ and $B$, whose outcome we want to estimate in terms of probabilities. 
The numbers of goals $G_A$ and  $G_B$ scored by teams $A$ and $B$ shall be random variables which follow a zero-inflated generalised Poisson-distribution (ZIGP). Generalised Poisson distributions generalise the Poisson distribution by adding a dispersion parameter; additionally, we add a point measure at $0$, since the event that no goal is scored by a team typically is a special event. We recall the definition that a discrete random variable $X$ follows a \textit{Zero-Inflated Generalized Poisson distribution (ZIGP)} with Poisson parameter $\mu>0$, dispersion parameter $\varphi\geq 1$ and zero-inflation $\omega\in[0,1)$:
$$
\mathbb{P}[X=k]=\begin{cases}
\omega + (1-\omega) \cdot e^{-\frac{\mu}{\varphi}}, & \textrm{if k=0,}\\
(1-\omega)\cdot \frac{\mu\cdot \bigl(\mu+(\varphi-1)\cdot k\bigr)^{k-1}}{k!}\varphi^{-k} e^{-\frac{1}{\varphi}\bigl(\mu+(\varphi-1)x\bigr)}, & \textrm{if $k\in\mathbb{N}$};
\end{cases}
$$
compare, e.g., with Consul \cite{Co:89} and Stekeler \cite{St:04}. If $\omega=0$ and $\varphi=1$, then we obtain just the classical Poisson distribution. The advantage of ZIGP is now that we have an additional dispersion parameter. We also note that
\begin{eqnarray*}
\mathbb{E}(X) &=& (1-\omega)\cdot \mu, \\
\mathrm{Var}(X) &=& (1-\omega)\cdot \mu \cdot (\varphi^2 + \omega \mu).
\end{eqnarray*}
The idea is now to model the number $G$ of scored goals of a team by a ZIGP distribution, whose parameters depend on the opponent's Elo ranking and the location of the match. Moreover, the number of goals scored by the weaker team (according to the Elo ranking) does additionally depend on the number of scored goals of the stronger team.
\par
We now explain the regression method in more detail.
In the following we will always assume that $A$ has \textit{higher} Elo score than $B$. This assumption can be justified, since usually the better team dominates the weaker team's tactics. Moreover the number of goals the stronger team scores has an impact on the number of goals of the weaker team. For example,  if team $A$ scores  $5$ goals it is more likely that $B$ scores also $1$ or $2$ goals, because the defense of team $A$ lacks in concentration  due to the expected victory. If the stronger team $A$ scores only $1$ goal, it is more likely that $B$ scores no or just one goal, since team $A$ focusses more on the defense  and tries to secure the victory.
\par
Denote by $G_A$ and $G_B$ the number of goals scored by teams $A$ and $B$. Both $G_A$ and $G_B$ shall be ZIGP-distributed: $G_A$ follows a ZIGP-distribution with parameter $\mu_{A|B}$, $\varphi_{A|B}$ and $\omega_{A|B}$, while $G_B$ follows a ZIGP-distribution with Poisson parameter $\mu_{B|A}$, $\varphi_{B|A}$ and $\omega_{B|A}$. These parameters are now determined as follows: 
\begin{enumerate}
\item In the first step we model the strength of team $A$ in terms of  the number of scored goals $\tilde G_{A}$   in dependence of the opponent's Elo score $\elo{}=\elo{B}$ and the location of the match. The location parameter $\mathrm{loc}_{A|B}$ is defined as: 
$$
\mathrm{loc}_{A|B}=\begin{cases}
1, & \textrm{if $A$ plays at home},\\
0, & \textrm{if the match takes place on neutral playground},\\
-1, & \textrm{if $B$ plays at home}.
\end{cases}
$$
The parameters of the distribution of $\tilde G_A$ are modelled as follows:
\begin{equation}\label{equ:independent-regression1}
\begin{array}{rcl}
\log \mu_A\bigl(\elo{B}\bigr)&=& \alpha_0^{(1)} + \alpha_1^{(1)} \cdot \elo{B} + \alpha_2^{(1)} \cdot \mathrm{loc}_{A|B},\\
\varphi_A &= &  1+e^{\beta^{(1)}}, \\
\omega_A &= & \frac{\gamma^{(1)}}{1+\gamma^{(1)}},
\end{array}
\end{equation}
where $\alpha_0^{(1)},\alpha_1^{(1)},\alpha_2^{(1)},\beta^{(1)},\gamma^{(1)}$  are obtained via ZIGP regression.  Here, $\tilde G_A$ is a model ofr the scored goals of team $A$, which does \textit{not} take into account the defense skills of team $B$.

\item Teams of similar Elo scores  may have different strengths in attack and defense. To take this effect into account  we model the  number $\check G_A$ of goals team $B$ receives  against a team of higher Elo score  $\elo{}=\elo{A}$ using a ZIGP distribution with mean parameter $\nu_{B}$, dispersion parameter $\psi_B$ and zero-inflation parameter $\delta_B$ as follows:
\begin{equation}\label{equ:independent-regression2}
\begin{array}{rcl}
\log \nu_B\bigl(\elo{A}\bigr) & =& \alpha_0^{(2)} + \alpha_1^{(2)} \cdot \elo{A} + \alpha_2^{(2)} \cdot \mathrm{loc}_{B|A},\\
\psi_B &= &  1+e^{\beta^{(2)}}, \\
\delta_B &= & \frac{\gamma^{(2)}}{1+\gamma^{(2)}},
\end{array}
\end{equation}
where $\alpha_0^{(2)},\alpha_1^{(2)},\alpha_2^{(2)},\beta^{(2)},\gamma^{(2)}$  are obtained via ZIGP regression. Here, we model the number of scored goals of team $A$ as the goals against $\check G_A$ of team $B$.

\item Team $A$  shall in average score $(1-\omega)\cdot \mu_{A}(\elo{B})$ goals against team $B$ (modelled by $\tilde G_A$), but team $B$ shall receive in average $(1-\omega_B)\cdot \nu_{B}(\elo{A})$ goals against (modelled by $\check G_A$). As these two values rarely coincides we model the numbers of goals $G_A$ as a ZIGP distribution with parameters
\begin{eqnarray*}
\mu_{A|B} &:= & \frac{\mu_A\bigl(\elo{B}\bigr)+\nu_B\bigl(\elo{A}\bigr)}{2},\\
\varphi_{A|B} &:=& \frac{\varphi_A + \psi_B}{2},\\
\omega_{A|B} &:= & \frac{\omega_A + \delta_B}{2}. 
\end{eqnarray*}

\item The number of goals $G_B$ scored by $B$ is assumed to depend on the Elo score $E_A=\elo{A}$, the location $\mathrm{loc}_{B|A}$ of the match and additionally on the outcome of $G_A$. Hence, we model $G_B$ via a ZIGP distribution with Poisson parameters $\mu_{B|A}$, dispersion $\varphi_{B|A}$ and zero inflation $\omega_{B|A}$ satisfying
\begin{equation}\label{equ:nested-regression1}
\begin{array}{rcl}
\log \mu_{B|A} &=& \alpha_0^{(3)} + \alpha_1^{(3)}  \cdot E_A+\alpha_0^{(3)}\cdot \mathrm{loc}_{B|A} +\alpha_3^{(3)}  \cdot G_A,\\
\varphi_{B|A} &:= &1+e^{\beta^{(3)}}, \\
\omega_{B|A} &:= & \frac{\gamma^{(3)}}{1+\gamma^{(3)}},
\end{array}
\end{equation}
where the parameters $\alpha_0^{(3)},\alpha_1^{(3)} ,\alpha_2^{(3)} ,\alpha_3^{(3)} ,\beta^{(3)} ,\gamma^{(3)} $ are obtained by ZIGP regression. 

\item The result of the match $A$ vs. $B$ is simulated by realizing $G_A$ first and then  realizing $G_B$ in dependence of the realization of $G_A$. 
\end{enumerate}

For a better understanding, we give an example and consider the match France vs. Germany, which takes place in Munich, Germany: France has $2087$ Elo points while Germany has $1936$ points. Against a team of Elo score $1936$ France is assumed to score without zero-inflation in average
$$
\mu_{\textrm{France}}(1936)=\exp\bigl(1.895766      -0.0007002232 \cdot 1936-         0.2361780 \cdot (-1)\bigr)=1.35521
$$
goals, and France's zero inflation is estimated as
$$
\omega_{\textrm{France}} = \frac{e^{-3.057658}}{1+e^{-3.057658}}=0.044888.
$$
Therefore, France is assumed to score in average
$$
(1-\omega_{\textrm{France}})\cdot \mu_{\textrm{France}}(1936) = 1.32516
$$
goals against Germany. Vice versa, Germany receives in average without zero-inflation
$$
\nu_{\textrm{Germany}}(2087)=\exp(-3.886702       +0.002203437   \cdot 2087      -0.02433679\cdot 1)=1.988806
$$
goals, and the zero-inflation of Germany's goals against is estimated as
$$
\delta_{\mathrm{Germany}} = \frac{e^{-5.519051}}{1+e^{-5.519051}}=0.003993638.
$$
Hence, in average Germany receives
$$
(1-\omega_{\textrm{Germany}})\cdot \nu_{\textrm{Germany}}(2087) = 1.980863
$$
goals against when playing against an opponent of Elo strength $2087$.
Therefore,   the number of goals, which France will score against Germany, will be modelled as a ZIGP distributed random variable with mean
$$
\Bigl(1-\frac{\omega_{\textrm{France}}+\delta_{\mathrm{Germany}}}{2}\Bigr)\cdot \frac{\mu_{\textrm{France}}(1936)+ \nu_{\textrm{Germany}}(2087)}{2}=1.627268.
$$

The average number of goals, which Germany scores against a team of Elo score $2087$ provided that $G_A$ goals against are received, is modelled by a ZIGP distributed random variable with parameters
$$
\mu_{\textrm{Germany}|\textrm{France}} = \exp\bigl( 3.340300      -0.0014539752\cdot 2087       -0.089635003  \cdot G_A   +    0.21633103\cdot 1\bigr);
$$
e.g., if $G_A=1$ then $\mu_{\textrm{Germany}|\textrm{France}}=1.54118$.
\par
As a final remark, we note that the presented dependent approach may also be justified through the definition of  conditional probabilities:
$$
\mathbb{P}[G_A=i,G_B=j] = \mathbb{P}[G_A=i]\cdot \mathbb{P}[G_B=j \mid G_A=i] \quad \forall i,j\in\mathbb{N}_0.
$$
For a comparision of this model in contrast to similar Poisson models, we refer once again to \cite{gilch:afc19} and \cite{gilch-mueller:18}. All calculations were performed with R (version 3.6.2). In particular, the presented model generalizes the models used in \cite{gilch-mueller:18} and \cite{gilch:afc19} by adding a dispersion parameter, zero-inflation and a regression approach which weights historical data according to importance and time depreciation. 
I

\subsection{Goodness of Fit Tests}\label{subsubsection:gof}
We check goodness of fit of  the ZIGP regressions in (\ref{equ:independent-regression1}) and (\ref{equ:independent-regression2})  for all participating teams. For each team $\mathbf{T}$ we calculate the following  $\chi^{2}$-statistic from the list of matches from the past:
$$
\chi_\mathbf{T} = \sum_{i=1}^{n_\mathbf{T}} \frac{(x_i-\hat\mu_i)^2}{\hat\mu_i},
$$
where $n_\mathbf{T}$ is the number of matches of team $\mathbf{T}$, $x_i$ is the number of scored goals of team $\mathbf{T}$ in match $i$ and $\hat\mu_i$ is the estimated ZIGP regression mean in dependence of the opponent's historical Elo points. 
\par
We observe that almost all teams have a very good fit. In Table \ref{table:godness-of-fit:gc} the $p$-values for some of the top teams are given. 
\begin{table}[H]
\centering
\begin{tabular}{|l|c|c|c|c|c|}
  \hline
 Team & Belgium & France & Portugal & Spain & Italy 
 \\ 
  \hline
  $p$-value & 0.98 &0.15 &  0.34 & 0.33  &0.93
    \\
   \hline
\end{tabular}
\caption{Goodness of fit test for the  ZIGP regression   in  (\ref{equ:independent-regression1}) for the top  teams. }
\label{table:godness-of-fit:gc}
\end{table}
Only Germany has a low $p$-value of  $0.05$; all other teams have a $p$-value of at least $0.14$, most have a much higher $p$-value.
\par

We  also calculate a $\chi^{2}$-statistic for each team which measures the goodness of fit for the regression in  (\ref{equ:independent-regression2}) which models the number of goals against. The $p$-values for the top teams are given in Table  \ref{table:godness-of-fit2:gc}.
\begin{table}[H]
\centering
\begin{tabular}{|l|c|c|c|c|c|}
  \hline
 Team & Netherlands & France & Germany & Spain & England
 \\ 
  \hline
  $p$-value & 0.26 &0.49 &  0.27 & 0.76  &0.29
    \\
   \hline
\end{tabular}
\caption{Goodness of fit test for the  ZIGP regression   in  (\ref{equ:independent-regression2}) for some of the top  teams. }
\label{table:godness-of-fit2:gc}
\end{table}
Let us remark that some countries have a very poor $p$-value like Italy or Portugal. However, the effect is rather limited since regression (\ref{equ:independent-regression2}) plays mainly a role for weaker teams by construction of our model.

\par

Finally, we test the goodness of fit for the regression in (\ref{equ:nested-regression1}) which models the number of goals against of the weaker team in dependence of the number of goals which are scored by the stronger team; see Table \ref{table:godness-of-fit3:gc}. 
\begin{table}[H]
\centering
\begin{tabular}{|l|c|c|c|c|c|}
  \hline
 Team &  Germany & England & Italy & Austria &  Denmark 
 \\ 
  \hline
  $p$-value  &0.06 &  0.41 & 0.91& 0.17 & 0.74
    \\
   \hline
\end{tabular}
\caption{Goodness of fit test for the  Poisson regression   in  (\ref{equ:nested-regression1}) for some of the  teams. }
\label{table:godness-of-fit3:gc}
\end{table}
Only Slovakia and Sweden have poor fits according to the $p$-value while the $p$-values of all other teams 
suggest reasonable fits.

\subsection{Validation of the Model}
\label{subsec:validation}

In this subsection we want to compare the predictions with the real result of the UEFA EURO 2016. For this purpose, we introduce the following notation: let $\mathbf{T}$ be a UEFA EURO 2016 participant. Then define:
$$
\mathrm{result}(\mathbf{T}) = \begin{cases}
1, &\textrm{if } \mathbf{T} \textrm{ was UEFA EURO 2016 winner}, \\
2, &\textrm{if } \mathbf{T} \textrm{ went to the final but didn't win the final},\\
3, &\textrm{if } \mathbf{T} \textrm{ went to the semifinal but didn't win the semifinal},\\
4, &\textrm{if } \mathbf{T} \textrm{ went to the quarterfinal but didn't win the quarterfinal},\\
5, &\textrm{if } \mathbf{T} \textrm{ went to the round of last 16 but didn't win this round},\\
6, &\textrm{if } \mathbf{T} \textrm{ went out of the tournament after the round robin}
\end{cases}
$$
E.g., $\mathrm{result(Portugal)}=1$, $\mathrm{result(Germany)}=2$, or $\mathrm{result(Austria)}=6$. For every UEFA EURO 2016 participant $\mathbf{T}$ we set the simulation result probability as $p_i(\mathbf{T}):=\mathbb{P}[\mathrm{result}(\mathbf{T})=i]$.
In order to compare the different simulation results with the reality we use the following distance functions:
\begin{enumerate}
\item \textbf{Maximum-Likelihood-Distance:} The error of team $\mathbf{T}$ is in this case defined as
$$
\mathrm{error}(\mathbf{T}) := \bigl| \mathrm{result}(\mathbf{T}) - \mathrm{argmax}_{j=1,\dots,6} p_i(\mathbf{T})]\bigr|.
$$
The total error score is then given by
$$
MDL = \sum_{\mathbf{T} \textrm{ UEFA EURO 2016 participant}} \mathrm{error}(\mathbf{T}) 
$$
\item \textbf{Brier Score:}
The error of team $\mathbf{T}$ is in this case defined as
$$
\mathrm{error}(\mathbf{T}) := \sum_{j=1}^6 \bigl( p_{j}(\mathbf{T}) -\mathds{1}_{[\mathrm{result}(\mathbf{T})=j]}  \bigr)^2.
$$
The total error score is then given by
$$
BS  = \sum_{\mathbf{T} \textrm{ UEFA EURO 2016 participant}} \mathrm{error}(\mathbf{T}) 
$$ 
\item \textbf{Rank-Probability-Score (RPS):} 
The error of team $\mathbf{T}$ is in this case defined as
$$
\mathrm{error}(\mathbf{T}) := \frac{1}{5} \sum_{i=1}^5 \left( \sum_{j=1}^i p_{j}(\mathbf{T}) -\mathds{1}_{[\mathrm{result}(\mathbf{T})=j]}  \right)^2.
$$
The total error score is then given by
$$
RPS  = \sum_{\mathbf{T} \textrm{ UEFA EURO 2016 participant}} \mathrm{error}(\mathbf{T}) 
$$ 
\end{enumerate}
We applied the model to the UEFA EURO 2016 tournament and compared the predictions with the basic Nested Poisson Regression model from \cite{gilch:afc19}.

\begin{table}[H]
\centering
\begin{tabular}{|l|c|c|}
  \hline
 Error function &  ZIGP & Nested Poisson Regression 
 \\ 
  \hline
   Maximum Likelihood Distance & 22 & 26 \\
  Brier Score & 17.52441& 18.68  \\
  Rank Probability Score &5.280199 & 5.36 
    \\
   \hline
\end{tabular}
\caption{Validation of ZIGP model compared with Nested Poisson Regression measured by different error functions. }
\label{table:godness-of-fit3:gc}
\end{table}

Hence, the presented ZIGP regression model seems to be a suitable improvement of the Nested Poisson Regression model introduced in \cite{gilch:afc19} and \cite{gilch-mueller:18}.

\section{UEFA EURO 2020 Forecast}

Finally, we come to the simulation of the UEFA EURO 2020, which allows us to answer the questions formulated in Section \ref{subsec:goals}. We simulate each single match of the UEFA EURO 2020 according to the model presented in Section \ref{sec:model}, which in turn allows us to simulate the whole UEFA EURO 2020 tournament. After each simulated match we update the Elo ranking according to the simulation results. This honours teams, which are in a good shape during a tournament and perform maybe better than expected.
Overall, we perform  $100.000$ simulations of the whole tournament, where we reset the Elo ranking at the beginning of each single tournament simulation. 

\section{Single Matches}

\begin{figure}[ht]
\begin{center}
\includegraphics[width=13cm]{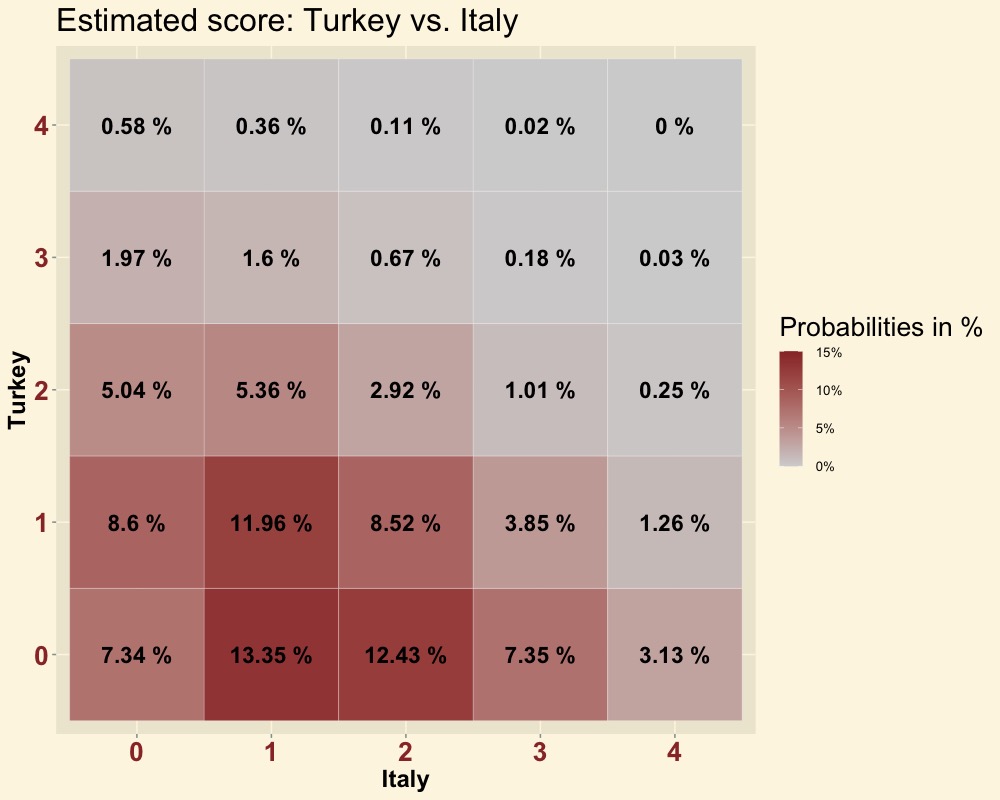}
\end{center}
\caption{Probabilities for the score of the match Turkey vs. Italy (group A) in Rome.}
\label{table:TR-IT}
\end{figure}

Since the basic element of our simulation is the simulation of single matches, we visualise how to quantify the results of single matches. Group A starts with the match between Turkey and Italy in Rome. According to our model we have the probabilities presented in Figure \ref{table:TR-IT} for the result of this match: the most probable scores are a $1:0$ or $2:0$ victory of Italy or a $1:1$ draw. 

\subsection{Group Forecast} 

In the following tables \ref{tab:EMgroupA}-\ref{tab:EMgroupF} we present the probabilities obtained from our simulation for the group stage, where we give the probabilities of winning the group, becoming runner-up, getting qualified for the round of last 16 as one of the best ranked group third (Third Q), 
or to be eliminated in the group stage. 
In Group F, the toughest group of all with world champion France, European champion Portugal and Germany, a head-to-head fight between these countries is expected for the first and second place.

\begin{table}[H]
\centering
\begin{tabular}{|rcccc|}
  \hline
 Team & GroupFirst & GroupSecond & Third Q & Prelim.Round \\ 
  \hline
Italy & 39.8 \% & 28.3 \% & 15.8 \% & 16.10  \%\\ 
 Switzerland & 24.1 \% & 26.9 \% & 19.6 \% & 29.40  \%\\ 
   Turkey & 23.7 \% & 25.5 \% & 19.1 \% & 31.80  \%\\ 
   Wales & 12.5 \% & 19.3 \% & 18.7 \% & 49.60  \%\\ 
   \hline
\end{tabular}
\caption{Probabilities for Group A}
\label{tab:EMgroupA}
\end{table}

\begin{table}[H]
\centering
\begin{tabular}{|rcccc|}
  \hline
 Team & GroupFirst & GroupSecond & Third Q & Prelim.Round \\ 
  \hline
 Belgium & 72.8 \% & 21.2 \% & 4.6 \% & 1.30  \%\\ 
   Denmark & 19.9 \% & 42.9 \% & 17.9 \% & 19.30  \%\\ 
Finland & 3.8 \% & 14.4 \% & 15.4 \% & 66.40  \%\\ 
  Russia & 3.5 \% & 21.4 \% & 18.4 \% & 56.70  \%\\ 
   \hline
\end{tabular}
\caption{Probabilities for Group B}
\label{tab:EMgroupB}
\end{table}

\begin{table}[H]
\centering
\begin{tabular}{|rcccc|}
  \hline
  Team & GroupFirst & GroupSecond & Third Q & Prelim.Round \\ 
  \hline
Netherlands & 57.8 \% & 27.5 \% & 9.7 \% & 5.00  \% \\ 
   Ukraine & 27.6 \% & 35.4 \% & 17.4 \% & 19.60  \%\\ 
   Austria & 10.6 \% & 24.9 \% & 24.9 \% & 39.70  \%\\ 
   North Macedonia & 4.1 \% & 12.3 \% & 14.5 \% & 69.20  \%\\ 
   \hline
\end{tabular}
\caption{Probabilities for Group C}
\label{tab:EMgroupC}
\end{table}

\begin{table}[H]
\centering
\begin{tabular}{|rcccc|}
  \hline
  Team & GroupFirst & GroupSecond & Third Q & Prelim.Round \\ 
  \hline
 England & 54.5 \% & 27.5 \% & 11.1 \% & 6.90  \%\\ 
   Croatia & 26.5 \% & 32.9 \% & 17.6 \% & 22.90  \%\\ 
   Czechia & 12.4 \% & 23.2 \% & 22.3 \% & 42.10  \%\\ 
   Scotland & 6.6 \% & 16.4 \% & 17.7 \% & 59.20  \%\\ 
   \hline
\end{tabular}
\caption{Probabilities for Group D}
\label{tab:EMgroupD}
\end{table}
  
\begin{table}[H]
\centering
\begin{tabular}{|rcccc|}
  \hline
  Team & GroupFirst & GroupSecond & Third Q & Prelim.Round \\ 
  \hline
 Spain & 71.9 \% & 19.8 \% & 5.9 \% & 2.40  \%\\ 
 Sweden & 12.6 \% & 34 \% & 20.6 \% & 32.80  \%\\   
 Poland & 12 \% & 31.6 \% & 21.6 \% & 34.90  \%\\ 
     Slovakia & 3.6 \% & 14.6 \% & 14.1 \% & 67.60  \%\\ 
   \hline
\end{tabular}  
\caption{Probabilities for Group E}
\label{tab:EMgroupE}
\end{table}

\begin{table}[H]
\centering
\begin{tabular}{|rcccc|}
  \hline
 Team & GroupFirst & GroupSecond & Third Q & Prelim.Round \\ 
  \hline
 France & 37.7 \% & 30.4 \% & 17.9 \% & 14.00  \%\\ 
 Germany & 32.4 \% & 30.3 \% & 19.9 \% & 17.40  \%\\ 
   Portugal & 26.4 \% & 29.9 \% & 23 \% & 20.60  \%\\ 
   Hungary & 3.5 \% & 9.5 \% & 11.8 \% & 75.10  \%\\ 
   \hline
\end{tabular}
\caption{Probabilities for Group F}
\label{tab:EMgroupF}
\end{table}

\subsection{Playoff Round Forecast}
Our simulations yield the following probabilities for each team to win the tournament or to reach certain stages of the tournament. The result is  presented in Table  \ref{tab:EURO2020}. The ZIGP regression model  favors Belgium, followed by the current world champions from France and Spain. The remaining teams have significantly less chances to win the UEFA EURO 2020.

\begin{table}[H]
\centering
\begin{tabular}{|rlllll|}
  \hline
 Team & Champion & Final & Semifinal & Quarterfinal & Last16 \\ 
  \hline
 Belgium & 18.4 \% & 29.1 \% & 47.7 \% & 68.7 \% & 98.5 \% \\ 
   France & 15.4 \% & 24.9 \% & 38.8 \% & 58.4 \% & 85.9 \% \\ 
   Spain & 13 \% & 22.5 \% & 38.9 \% & 68.1 \% & 97.7 \% \\ 
   England & 7.8 \% & 14.8 \% & 26.8 \% & 50.5 \% & 93 \% \\ 
   Portugal & 7.7 \% & 15.5 \% & 28.6 \% & 47.7 \% & 79.5 \% \\ 
   Netherlands & 7.1 \% & 14.7 \% & 28.9 \% & 54.4 \% & 95 \% \\ 
   Germany & 6.1 \% & 13.6 \% & 27.4 \% & 47.3 \% & 82.6 \% \\ 
   Italy & 4.8 \% & 10.7 \% & 22.7 \% & 47.7 \% & 83.9 \% \\ 
   Turkey & 3.7 \% & 8 \% & 17.1 \% & 35.2 \% & 68.2 \% \\ 
   Denmark & 3.5 \% & 9.2 \% & 20.8 \% & 40.7 \% & 79.7 \% \\ 
   Croatia & 3.4 \% & 8.3 \% & 17.7 \% & 38.1 \% & 76.9 \% \\ 
   Switzerland & 3.2 \% & 7.9 \% & 17.9 \% & 37.6 \% & 70.6 \% \\ 
   Ukraine & 1.8 \% & 5.1 \% & 13.2 \% & 34.4 \% & 80.4 \% \\ 
   Poland & 1.2 \% & 3.9 \% & 10.9 \% & 29.6 \% & 66.6 \% \\ 
   Sweden & 1.1 \% & 3.9 \% & 10.8 \% & 30.9 \% & 68.6 \% \\ 
   Wales & 0.6 \% & 2.2 \% & 7.2 \% & 19.7 \% & 50.6 \% \\ 
   Czechia & 0.4 \% & 1.7 \% & 5.9 \% & 19.6 \% & 57.9 \% \\ 
   Russia & 0.2 \% & 1 \% & 4.1 \% & 12.9 \% & 41.5 \% \\ 
   Finland & 0.1 \% & 0.7 \% & 2.6 \% & 9 \% & 32.3 \% \\ 
   Austria & 0.1 \% & 0.6 \% & 3.3 \% & 15 \% & 60.3 \% \\ 
   Slovakia & 0.1 \% & 0.7 \% & 2.7 \% & 10.1 \% & 34 \% \\ 
   Hungary & 0.1 \% & 0.6 \% & 2.6 \% & 8.3 \% & 24.7 \% \\ 
   Scotland & 0.1 \% & 0.5 \% & 2.3 \% & 10.6 \% & 40.7 \% \\ 
   North Macedonia & 0 \% & 0.1 \% & 0.9 \% & 5.5 \% & 30.8 \% \\ 
   \hline
\end{tabular}\caption{UEFA EURO 2020 simulation results for the teams' probabilities to proceed to a certain stage}
\label{tab:EURO2020}
\end{table}


%
%

\section{Final remarks}
\label{sec:discussion}

As we have shown in Subsection \ref{subsec:validation} the proposed ZIGP model with weighted historical data seems to improve the model which was applied in  \cite{gilch:afc19} for CAF Africa Cup of Nations 2019.
For further discussion on adaptions and different models, we refer once again to the discussion section in \cite{gilch-mueller:18} and \cite{gilch:afc19}.

%
%

\bibliographystyle{apalike}
\bibliography{bib}

\end{document}